\documentclass[11pt,twoside]{article}
\usepackage{./asp2014}
\usepackage{graphicx}
\usepackage{wrapfig}
\usepackage{lscape}
\usepackage{rotating}
\usepackage{epstopdf}

\aspSuppressVolSlug
\resetcounters

\begin{document}

\title{Imaging Stellar Radio Photospheres with the Next Generation Very Large Array}

\author{C.L. Carilli,$^1$ B. Butler,$^1$ K. Golap,$^1$, M.T. Carilli,$^2$ and S.M. White$^4$}
\affil{$^1$ National Radio Astronomy Observatory, Charlottesville, VA, 22903; \email{ccarilli@nrao.edu}}
\affil{$^2$ University of Colorado, Boulder, CO}
\affil{$^3$ Air Force Research Lab, Albuquerque, NM}

\paperauthor{C.L. Carilli}{ccarilli@nrao.edu}{}{National Radio Astronomy Observatory}{}{Socorro}{NM}{87801}{USA}
\paperauthor{B. Butler}{}{}{National Radio Astronomy Observatory}{}{Socorro}{NM}{87801}{USA}
\paperauthor{K. Golap}{}{}{National Radio Astronomy Observatory}{}{Socorro}{NM}{87801}{USA}
\paperauthor{M.T. Carilli}{}{}{University of Colorado}{}{Boulder}{CO}{80309}{USA}
\paperauthor{S. White}{}{}{AFRL}{Kirtland Airforce Base}{Albuquerque}{NM}{}{USA}


\begin{abstract}
We perform simulations of the capabilities of the Next Generation Very
Large Array to image stellar radio photospheres. For very large (in
angle) stars, such as red supergiants within a few hundred parsecs,
good imaging fidelity results can be obtained on radio photospheric
structures at 38\,GHz employing standard techniques, such as disk model
fitting and subtraction, with hundreds of resolution elements over
the star, even with just the ngVLA-classic baselines to 1000\,km. Using the
ngVLA Rev B plus long baseline configuration (with baselines
out to 9000\,km, August 2018), we find
for main sequence stars within $\sim 10$\,pc, the photospheres can be
easily resolved at 85\,GHz, with accurate measures of the mean
brightness and size, and possibly imaging large surface structures,
as might occur on eg. active M dwarf stars. 
For more distant main sequence stars, we find
that measurements of sizes and brightnesses can be made using disk
model fitting to the $u,v$ data down to stellar diameters $\sim 0.4$\,mas in a
few hours. This size would include M0 V stars to a distance of 15\,pc,
A0 V stars to 60\,pc, and Red Giants to 2.4\,kpc. Based on the Hipparcos
catalog, we estimate that there are at least 10,000 stars that will be
resolved by the ngVLA. While the vast majority of these (95\%) are giants
or supergiants, there are still over 500 main sequence stars that can be
resolved, with $\sim 50$ to 150 in each spectral type (besides O stars).
Note that these are lower limits, since radio photospheres can be
larger than optical, and the Hipparcos catalog might not be complete.
Our initial look into the Gaia catalog suggests these numbers might
be pessimistic by a factor few. 

\end{abstract}

\section{Introduction}

The field of stellar atmospheres, and atmospheric activity, has taken
on new relevance in the context of the search for habitable planets,
due to the realization of the dramatic effect 'space weather' can have
on the development of life \citep{ost18, trig18}.  Observations at
radio wavelengths provide some of the most powerful probes of
exo-space weather, including: (i) high brightness coherent processes,
such as masers associated with outflows in forming stars, (ii)
non-thermal coherent and incoherent gyro-synchrotron emission due to
magnetic activity in stellar coronae, and (iii) thermal emission from
extreme mass loss events, such as novae \citep{gud02, lin18}.

In this article, we consider the more quiescent properties of stellar
radio photospheres, namely, the thermal emission from the $\tau \sim
1$ surface of the photospheres of stars of different types. The size
of stars, as measured by their stellar radio photospheres, is a strong
function of wavelength, since the free-free optical depth drops
quickly with frequency ($\propto \nu^{-2.1}$). Hence, radio
observations over a range in frequencies provide a tomographic image
of the density and temperature structure of the transition from the
optical photosphere to the chromosphere. For giant stars, the radio
photosphere will be larger, and typically cooler, than the optical
photosphere at long wavelengths, and approach the optical photospheric
size and temperature at shorter wavelengths. For instance, in red
super-giants, the radio photospheres at $\sim 6$cm are typically
$7\times$ larger, and $3\times$ cooler, than the optical photosphere,
while at $\sim 1$mm the radio and optical photospheric temperatures
and sizes are comparable to within $\le 30\%$ \citep{lim98, ogor15,
ogor17, har13}. For main sequence stars, there should be closer
correspondance between radio and optical photospheric sizes, although
even in these cases, at cm and mm wavelengths the stellar atmospheres
can become optically thick in the chromosphere, which would enhance
temperatures by a few thousand Kelvin (eg. the Sun has a brightness
temperature at 100\,GHz of 7300\,K\citep{white17}).

Of particular interest is the question of the end-of-life mass loss in
red giants and supergiants through strong winds. Current observations
suggest elevated atmospheres, with substantial large-scale temperature
structure, heated by sound or Alfven waves driven by large convective
cells bubbling-up from the stellar interior \citep{lim98, ogor15,
ogor17, har18, MC18}. For Solar-type and lower
mass main sequence stars, recent interest has focused on atmospheric
conditions that can drive strong winds, leading to star-planet
magnetic interactions, possibly triggering auroral radio emission
\citep{trig18, fich17, hal15}.

To date, direct imaging of stellar radio photospheres has been limited
to a handful of the largest, closest stars, such as red giants,
super-giants, and AGB stars \citep{lim98, vlem17, ogor13, MC15}. 
Likewise, detection of thermal emission from nearby
solar-type main sequence stars is just possible with deep observations
with the JVLA and ATCA \citep{vill14, ogor17}, but imaging is
certainly impossible with current facilities. The situation is similar
in the optical and near-IR, with a moderate number of large, hot stars
resolved through optical interferometry or direct imaging with HST
\citep{boy15, monn07, gil96, moz03}. For optical interferometers,
structure is typically inferred through model fitting of
interferometric closure quantities \citep{monn03, defr18}, although
hybrid closure-imaging approaches are being explored with the new VLTi
\citep{mont18}.

The `Next Generation Very Large Array' (ngVLA) opens a new window on
stellar radio photospheres, through sub-milliarcsecond resolution at
up to 90\,GHz with sub-$\mu$Jy sensitivity \citep{mck16, car15, sel17}.
The ngVLA rms brightness sensitivity in one hour at 85\,GHz and 0.4\,mas
resolution is $\sim 1000$\,K. There are thousands of stars with sizes
larger than 0.4\,mas, with brightness temperatures above a few thousand
Kelvin at 90\,GHz.  These include nearby main sequence stars, and more
distant giants and supergiants.  We investigate the ability of the
ngVLA to provide high fidelity images of the radio photospheres of
larger (in angle) stars, and to determine sizes and flux densities for
sub-mas stars.

\section{Configuration, Model, and Process}

We employ the newly accepted ngVLA configuration (August 2018). This
configuration includes the original ngVLA Rev B configuration
distributed across the US Southwest and Mexico. The Rev B
configuration includes 214 antennas of 18 meter diameter, to baselines up
to 1000\,km \citep{carer18, sel17}. The new ngVLA configuration
includes 30 more antennas distributed to baselines of 9000\,km,
including clusters of two to three antennas at sites such as Hawaii
and St. Croix.

We employ the {\it SIMOBSERVE} task within CASA to generate $u,v$ data sets.
Instructions on how this is done are found on the ngVLA web page.  

The ngVLA has a highly centrally condensed antenna distribution. The
naturally weighted beam for this centrally condensed distribution
leads to a very non-Gaussian PSF, with two skirts extending to
$10\times$ the nominal full resolution of the longest baselines at the
50\% to 20\% level, and $100\times$ the full resolution at the 20\% to
10\% level, for just the Rev B array alone.  This PSF problem becomes
an order of magnitude more severe when including the longer baselines,
in terms of the relative angular size of the PSF 'spike' corresponding
to the longest baselines vs. the skirts.  These prominent skirts
represent the major challenge when trying to image complex
structure. The imaging process entails a balance between sensitivity
and synthesized beam shape, using visibility tapering and Briggs
robust weighting \citep{cjs18, cot17, car17, car16}.  In all cases we
employed the {\it CLEAN} algorithm with Briggs weighting, and adjusted
the robust parameter and $u,v$-taper to give a reasonable synthesized beam
and noise performance.

We employ a template stellar disk model that includes a few large cold
and hot spots at the $\pm 10\%$ level (Fig 1a) .  Such spots have been
seen on the radio photospheres of giant and supergiant stars.  Such
large structures are not present on most main sequence stars, but they
provide some 'test structures' that exercise the imaging alogrithms at
high resolution. Moreover, active stars, such as M-dwarves, may have large
temperature structures, such as giant sunspot regions. We then adjust
the star size and mean brightness temperature according to the stellar
types and assumed distance.  We emphasize that the models below are
not meant to be accurate reproductions of a specific star, just
representative.

The new long baselines extend from 1000\,km to 9000\,km. The longest
baselines that include correlation to the ngVLA core are about 5000\,km
(eg. Socorro to Hawaii). Beyond that, only outer stations are involved
in the cross correlation, and we will show that the sensitivity drops
dramatically beyond 5000\,km baselines, as expected.

\begin{figure}[]
\begin{center}
\includegraphics[scale=0.6,angle=-90]{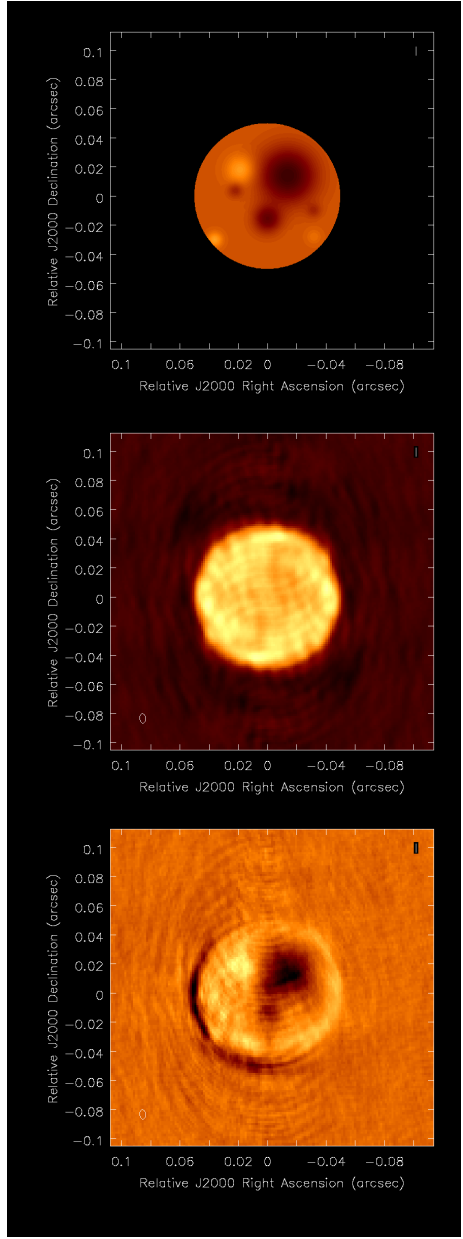}
\end{center}
\caption{\footnotesize {Images of Betelgeuse at 38\,GHz.
Top: the input model, including cold and hot spots
of peak magnitude $\sim \pm 12\%$. 
Center: The ngVLA image at 6.5mas $\times$ 4.0mas resolution,
for a 4 hour synthesis.  The rms noise is 0.51\,$\mu$Jy\,beam$^{-1}$,
and the peak surface brightness is 0.1 mJy\,beam$^{-1}$.
Bottom: Image with the ngVLA, 
after fitting for a disk model, and subtracting the model
from the visibilities. The rms noise is 0.25\,$\mu$Jy\,beam$^{-1}$,
and the peak surface brightness is 7.8\,$\mu$Jy\,beam$^{-1}$.
}}
\end{figure}

\section{Results}

\subsection{Betegeuse ($\alpha$ Orionis) at 38\,GHz}

Betelgeuse ($\alpha$ Orionis, J0555+0724, spectral type M1 Ia, Mass =
11.7 M$_\odot$), is a red supergiant at a distance of 222\,pc.  For the
Betelgeuse-like model, we adopt parameters for the radio photosphere
guided by existing observations at 22\,GHz to 43\,GHz from the VLA
\citep{lim98, ogor17}. The model has a diameter or 100\,mas and a total
flux density of 21.7 mJy at 1 cm wavelength. The implied mean
brightness temperature is about 3000\,K. The model includes large cold
and hot spots with Gaussian profiles, with sizes (FWHM) up to 20\% of
the stellar diameter, and peak deviations from the mean of $\pm 12\%$,
qualitatively consistent with radio imaging of Betelgeuse.

The model is shown in Figure 1a. We simulated a 4 hour observation of
this model with the ngVLA at 38\,GHz (Figure 1b). For imaging, we adopt
Briggs weighting with robust = $-0.5$, and we use a multiscale clean
with a loop gain of 0.03 to mitigate the CLEAN instability.  The
resulting image has an rms noise of 0.6\,$\mu$Jy\,beam$^{-1}$, and the
synthesized beam has a FWHM = 6.5\,mas $\times$ 4.0\,mas. The disk is
imaged at high signal to noise, however, the hot and cold spots are
only marginally perceptible. There is a depression at the position of the
largest cold spot of $\sim 15\%$, but the overall impression is that
of an edge-brightened disk.  This result is a well known issue with
interferometric imaging of objects with relatively flat brightness
distributions and hard edges, such as the disks of planets in our
Solar system \citep{depat16}. The result is due to the natural tendancy for an
interferometer to enhance edges and sharp gradients in images, and the
general CLEAN instability for diffuse sources \citep{corn08}.

In the case of Solar system planets, the standard analysis technique
is to perform disk model fitting to the visibilities, then subtract
the disk model from the $u,v$ data, to look for under-lying structures
\citep{depat16}.  We have performed such an analysis on the Betelgeuse
simulation.  The results from the $u,v$ -model fit are listed in Table 1.
The residual image after model subtraction from the visibilities is
shown in Figure 1c. The model fit provides an accurate realization of
the mean disk, both in terms of size and total flux density, to better
than 1\%. The residual image shows the more prominent hot and cold
spots, with magnitudes within 20\% of the expected deviations. One
artifact of the model and fitting process is that the largest cold
spot is significant enough to perturb the central position of the fit.
Even a $\sim 0.5\%$ position shift (relative to the stellar diameter),
will lead to sharp positive/negative edges in the residual image.

\begin{table}
\centering
\footnotesize
\caption{$u,v$ Fitting Results}
\begin{tabular}{lccccc}
\hline
Star & Max Baseline & S$_{mod}$ & D$_{mod}$ & S$_{fit}$ & D$_{fit}$ \\
~ & $\times 10^8 \lambda$ & mJy & mas & mJy & mas \\
\hline
$\alpha$ Orionis 38\,GHz & 1.3 & 21.7 & 100 & $21.6\pm 0.2$ &  $100.6 \pm 0.5$ \\
$\alpha$ Can Maj 85\,GHz & 3.0 & 1.63 & 6.0 
& $1.6295\pm 0.0002$ &  $6.039 \pm 0.005 \times 6.042\pm0.003$  \\
$\alpha$ Can Maj 85\,GHz & -- & 1.63 & 6.0
& $1.6295\pm 0.0002$ &  $6.040 \pm 0.005 \times 6.043\pm0.003$  \\
$\theta$ Leonis 85\,GHz  & 3.0 & 0.0255 & 0.723
& $0.0251\pm 0.00077$ &  $0.58\pm0.41 \times 0.61\pm0.12$ \\
$\theta$ Leonis 85\,GHz & -- & 0.0255 & 0.723
& $0.0251\pm 0.00075$ &  $0.56\pm0.17 \times 0.63\pm0.08$ \\
A2 V at 100\,pc 85\,GHz & -- & 0.0064 & 0.36 & $0.0063\pm 0.0004$
&  $0.45\pm 0.14\times 0.33\pm 0.10$ \\
\hline
\vspace{0.1cm}
\end{tabular}
\end{table}

\begin{figure}[]
\includegraphics[scale=0.6,angle=-90]{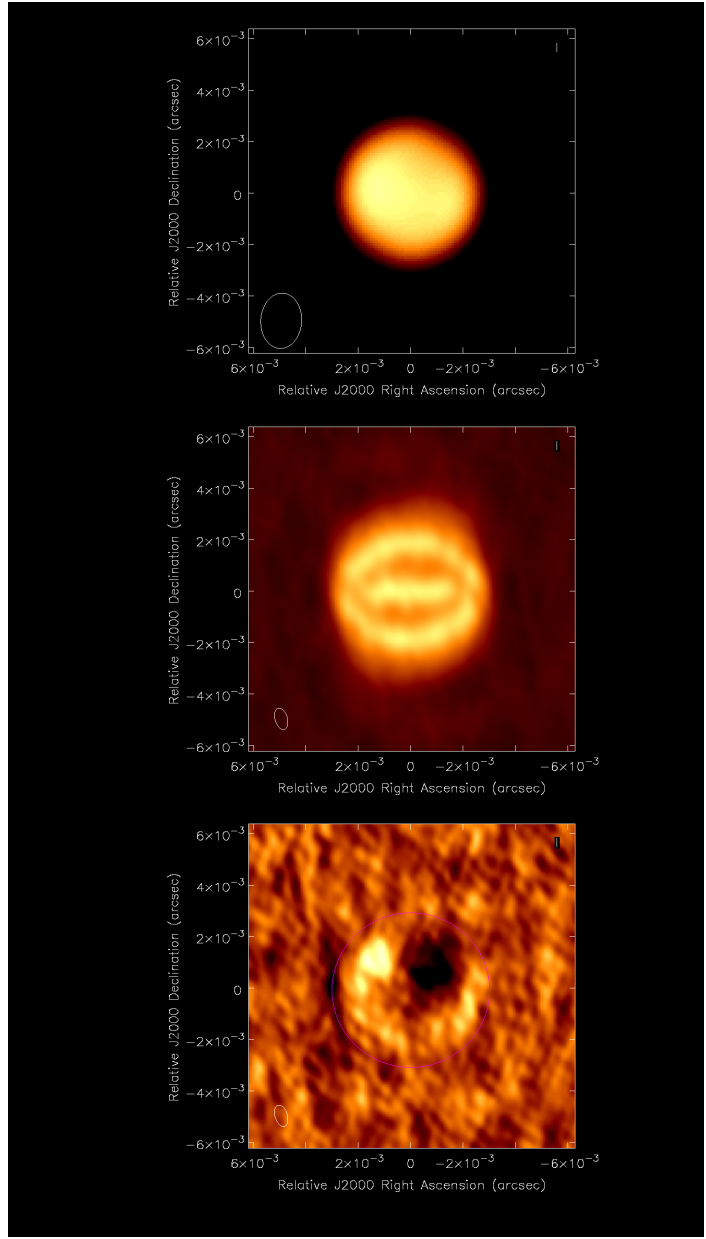}
\caption{\footnotesize {Top:   Image of Sirius with the
ngVLA for a 16 hour synthesis, made using $u,v$-weighting appropriate for
the original Rev B configuration (1000~km maximum baseline). 
The resolution is 2.1mas $\times$ 1.6mas and the rms = 0.6\,$\mu$Jy\,beam$^{-1}$.
Middle: Image of Sirius with $u,v$-weighting that
incorporates the longer baselines.
The resolution is 0.85mas $\times$ 0.47mas and the
rms = 0.5\,$\mu$Jy\,beam$^{-1}$.
Bottom: Image of Sirius with the same resolution as (b), 
after fitting for a disk model and subtracting 
the model from the $u,v$ data.
The pink circle shows roughly the position of the stellar disk.
The peak on the image is 3.5$\mu$Jy\,beam$^{-1}$. (Note: please refer
to Figure 1a for the model structure, scaled accordingly). 
}}
\end{figure}

\subsection{Sirius ($\alpha$ Canis Majoris) at 85\,GHz}

Sirius is the brightest star in the sky in the optical ($\alpha$ Canis
Majoris, J0645-1642), at a distance of 2.6\,pc from Earth. It is a
binary system, with the brighter star being a hot, 2 M$_\odot$, main
sequence star of spectral type A0, and the second star being a 1
M$_\odot$ white dwarf. We focus on the A0 star.

For the radio model of a Sirius-like star, and subsequent analyses
below, we adopt the parameters of the optical photosphere, in terms of
diameter and brightness temperature, since the radio and optical
photospheres for hot main sequence stars at 85\,GHz are likely to be
similar at the $\sim 30\%$ level, in size and brightness
\citep{white18}. We feel this assumption is conservative, in that the
radio photospheres could be larger.
In this model, the angular diameter is 6\,mas and the
total flux density at 85\,GHz is 1.63 mJy. The mean brightness
temperature is about 9000\,K.  We incorporate the same stellar structure as
in Section 3.1 (hot and cold spots).  Such structures are unlikely in
hot main sequence stars, but they allow us to explore the imaging
capabilities of the array at the highest resolutions.

We simulate a 16 hour observation at 85\,GHz with the Rev B
configuration plus the long baselines. We first image with a taper
appropriate to the scales of the Rev B baselines only: $R = -1.3$, a
cell size of 0.1\,mas, and an outer $u,v$-taper of 1.4\,mas. The resulting
image is shown in Figure 2a. The beam FWHM is 2.1\,mas $\times$ 1.6\,mas,
and the rms is 0.6\,$\mu$Jy\,beam$^{-1}$ (the Naturally weighted
theoretical noise is 0.2\,$\mu$Jy\,beam$^{-1}$). The resulting image
shows a resolved star, but the resolution is inadquate to isolate any
region clearly. There may be a depression at the location of the
biggest cold spot, but again, the image is not conclusive.

We then image with a taper more appropriate for the longer baselines:
$R = -1.3$, cell size of 0.05\,mas, and an outer $u,v$-taper of 0.35\,mas. The
resulting synthesized beam FWHM is 0.85\,mas $\times$ 0.47\,mas, and the
image noise is about 0.5\,$\mu$Jy\,beam$^{-1}$.  Figure 2b shows the
resulting image.  The structure is clearly dominated by imaging
artifacts.

We perform a similar disk $u,v$-model fitting and subtraction as in Section
3.1. The resulting image of the residuals is show in Figure 2c,
using the same weighting as is used in 2b.
There is a reasonable indication of both
the primary cold and hot spots, at levels within 30\% of the expected
values relative to the mean disk brightness, and even marginal
evidence for the next lowest cold spot.

The results for the disk fitting to the visibilities are given in
Table 1. We performed two fits: one using baselines from the
original Rev B configuration out to 1000\,km (3e8 $\times \lambda$),
and the second using all the long baselines (out to 9000\,km).
In this case, the results are much the same, since the star is 6\,mas
in diameter, and the resolution of even the Rev B baselines is
a factor few smaller than this. 

\begin{figure}[]
\begin{center}
\includegraphics[scale=0.5,angle=-90]{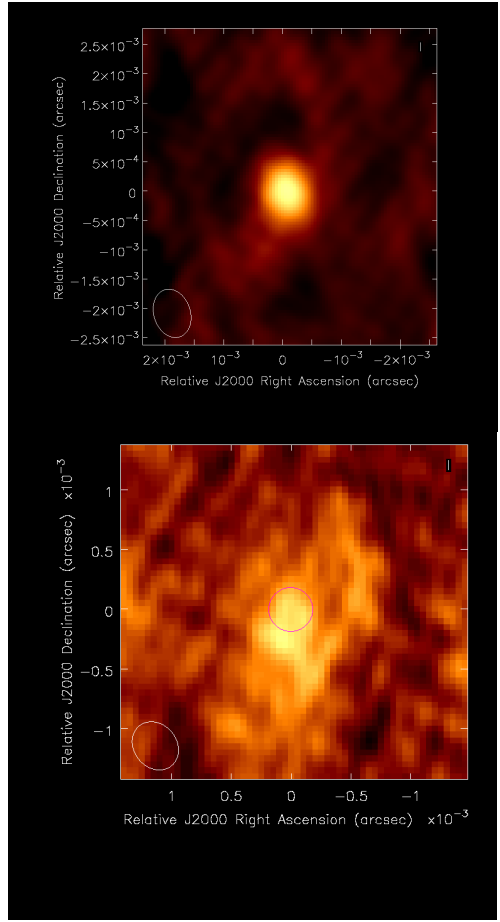}
\end{center}
\caption{\footnotesize {Top: Image of
$\theta$ Leonis with the Rev B plus long
baseline configuration of August, 2018, for a one hour synthesis. 
The resolution is 0.84\,mas $\times$ 0.62\,mas and the rms = 2\,$\mu$Jy\,beam$^{-1}$.
The peak on the image is 25\,$\mu$Jy\,beam$^{-1}$.
Bottom: Image of an A2 V star at a distance of 10\,pc using
the Rev B plus long baseline configuration of August, 2018,
for a 4 hour synthesis. 
The resolution is 0.43\,mas $\times$ 0.36\,mas and the
rms = 0.7\,$\mu$Jy\,beam$^{-1}$.
The peak in the field is 3.6\,$\mu$Jy\,beam$^{-1}$.
The circle shows the model photospher position and size. 
}}
\end{figure}

\subsection{$\theta$ Leonis at 85\,GHz}

We next consider a model of a hot main sequence star at a distance 20
times that of Sirius, and determine how well the stellar properties
can be determined in a one hour synthesis, appropriate for surveys of
large numbers of stars.  We adopt the parameters comparable to those
of $\theta$ Leonis, a 2.5 M$_\odot$ A2 V star at a distance of 51\,pc
(J1114+1525). The model in this case has a total flux density at 85\,GHz of 25.5\,$\mu$Jy, and a diameter of 0.723\,mas. We simulate a 1 hour
observation at 85\,GHz using the Rev B plus long baseline
configuration. The resulting image is shown in Figure 3. We use $R =
-1.0$ and an outer taper of 0.35\,mas. The beam FWHM = 0.84\,mas
$\times$ 0.62\,mas, and the noise is about 2.0\,$\mu$Jy\,beam$^{-1}$. The
resulting image is shown in Figure 3a.

\begin{figure}[]
\begin{center}
\includegraphics[scale=0.7]{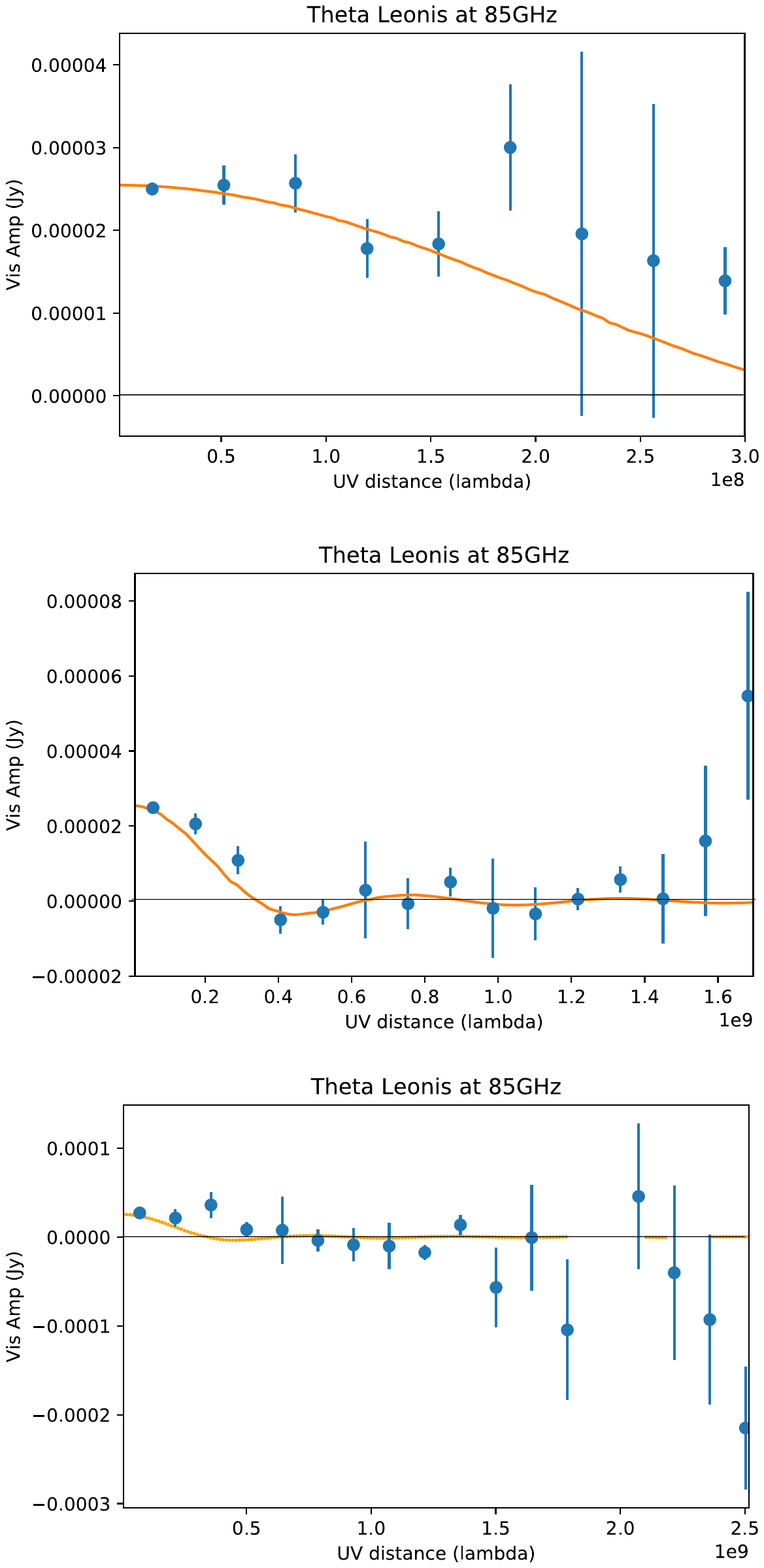}
\end{center}
\caption{\footnotesize{Azimuthally averaged Real part of the visibility 
curve for $\theta$ Leonis at 85\,GHz. Points are the measurements for a 4 hour 
synthesis with the ngVLA-classic baselines to 1000\,km (top)
and the Rev B plus long baseline configuration truncated at about
5000\,km (middle), and at 9000\,km (bottom).
Beyond about 5000\,km, the measurements become very noisy since the
large number of antennas in the Core and Plains no longer contribute. 
}}
\end{figure}

More relevant than direct imaging in this case is $u,v$ model
fitting. Table 1 shows disk modeling fitting results, again using
baselines appropriate for the Rev B configuration only, and then
including the long baselines. In this case, the total flux
density is well fit by both short and long baseline fits,
but the errors on the diameters are significantly lower
when including the longer baselines. 

Figure 4 shows a radial binning of the Real part of the visibilities
vs $u,v$-distance, plus the model visibilities. The Rev B only
configuration does not reach the first null in the visibility
function. The longer baselines sample the first three nulls at
reasonable signal to noise, out to $1.4\times 10^9$ wavelengths (=
5000\,km).  Beyond that, the sensitivity drops dramatically. This
sensitivity drop is due to the fact that only the outer antennas are
involved in the cross correlation, and the Core and Plains antennas no
longer contribute.

\subsection{An A2 V Star at 100 pc: limiting case}

To test the limits of the array, we model a hot main
sequence star as above (A2 V star), but now at a distance of 100\,pc,
or 40 times further than Sirius. We simulate a 4 hour synthesis
at 85\,GHz. The total flux density in the model is 6.4\,$\mu$Jy, and the stellar
diameter is 0.36\,mas. We make an image using R = -1.0, taper = 0.15\,mas,
and a cell size of 0.05\,mas. The resulting resolution is 0.4\,mas and
rms noise is 0.70\,$\mu$Jy (vs. 0.36\,$\mu$Jy Naturally weighted theoretical noise).
The resulting image is shown in Fig 3b.
There is a fuzzy source at the expected position, but determining the
size and total flux density from such an image is problematic, and
disk fitting to the visibilities is the only viable method. 

The results for disk fitting to the visibilities are
again given in Table 1 using the full array.
We find that the flux density is well determined ($\ge 10\sigma$),
and the diameter is fit to $\sim 3\sigma$.  We consider this the limiting
case for determining stellar properties with the ngVLA. 

\begin{table}
 \centering
\caption{Hipparcos Stars $> 0.4$ mas, $\rm Dec >-40^o$}
\begin{tabular}{ccccccc}
\hline
 ~ & I & II & III & IV & V & Total \\
O   &   6 & 1    & 0   & 0   & 2   & 9 \\
B   & 114 &   77 &   81 &   55 &   61 &  388 \\
A   &  35 &   32 &  105 &  118 &   61 &  351 \\
F   &  55 &   47 &  148 &  245 &  159 &  654 \\
G   &  67 &  106 & 1120 &  351 &  107 & 1751 \\
K   &  55 &  117 & 5351 &  227 &  104 & 5854 \\
M   &  27 &   32 & 1039 &   2 &   44 & 1144 \\
Total & 359 &  412 & 7884 &  998 &  538 & 10151 \\
\hline
\vspace{0.1cm}
\end{tabular}
\end{table}

\subsection{Number of Stars}

Based on the simulations above, we adopt an extreme limiting case for
determining stellar parameters of: a $1\sigma$ limit in 1hr
of $1.4~ \mu$Jy\,beam$^{-1}$ at 85\,GHz and 0.4\,mas stellar size. 
This limit imples an rms brightness temperature
limit of 1500\,K.  M-type stars have optical photospheric
temperatures of $\sim 3000$ to
4000\,K, which would be at the limit of detection at this resolution.
However, as mentioned above, at cm and mm wavelengths, the stellar
atmospheres of main sequence stars
can become optically thick in the chromosphere, which
would enhance temperatures by a few thousand Kelvin. 
A size of 0.4\,mas corresponds to the diameter of an M0 V star
(optical radius of $0.6 R_{\odot}$) at a distance of 15\,pc.  For A0 V
stars (radius of $2.5 R_{\odot}$), the distance is 60\,pc. Red giant
stars (radii of order $100 R_{\odot}$), could be resolved to 2.5\,kpc.

The Hipparcos catalog has a detailed listing of stellar spectral type,
luminosity class, and distance \citep{per97}.  From these quantities,
we derive angular sizes using standard relationships for stellar
optical photospheric diameters from Allen's Astrophysical Quantities
(Table 15.8). We then select for stars above $-40^o$ Dec, with
angular diameters $> 0.4$\,mas. The results as a function of spectral
type and luminosity class, are given in Table 2. The total number of
stars evaluated was 32886 (meaning stars with fully specified spectral
type and luminosity class, not Carbon stars, and above $-40^o$
Dec\footnote{The 32886 was out of a total number of stars in the
catalog of 118218, of which 91488 are above $-40^o$ Dec}). Of these,
10151 are larger than 0.4\,mas. The vast majority of the resolved stars
are in the Giant or Supergiant class (9613 or 95\% in class IV or
lower). Yet there are still more than 500 main sequence stars (class
V), that could be resolved by the ngVLA, with $\sim$ 50 to 150 in each
spectral type, except for the O stars.

For those interested in making direct comparisons to the Sun, there
are 43 'Solar-type' stars ($B-V$ color index between 0.65 and 0.67),
and 145 'Solar-analog' stars ($B-V$ color index between 0.62 and
0.71), that could be resolved by the ngVLA. 

We note that these are lower limits, since the radio photospheres can
be substantially larger than the optical photospheres, in particular
in Giants and Supergiants, or hotter, in the case of main sequence
stars. Additionally, the Hipparcos catalog is much smaller than the new GAIA
survey \citep{and18}. However, by definition, the larger (in angle)
stars of a given type/class will also be closer, and hence brighter
and more likely to be in both catalogs.

For completeness, we have also searched the Gaia catalog. Of the
almost 1.7 billion stars in that catalog, there are 77 million which
have both effective temperature and radius cataloged, of which 54
million are above $-40^o$ Dec. Based on the radii and distances in the
catalog, we find that there are about 40,000 stars larger than 0.4\,mas
and Dec $\ge -40^o$. Of these, we estimate that 246 are ``solar analogs",
meaning with tabulated radius and temperature within 30\% Solar.  The
Gaia catalog does not contain information about spectral type or
luminosity class, so no table similar to Table 2 for the Hipparcos
catalog can be made, nor does it contain values for $B-V$, so we cannot
define 'solar-type' stars in quite the same way as for Hipparcos.

\section{Conclusions}

The ngVLA will transform the field of stellar radio photospheres,
allowing for imaging and parameter estimation of thousands of
main sequence and giant stars.  Such studies are directly relevant to
determining the structure of stars in the key transition region from
the optical photosphere to the chromosphere -- the regions powering
exo-space weather phenomena.

For very large (in angle) stars, such as red supergiants within a few
hundred parsecs, good fidelity results can be obtained, with many
resolution elements across the star. Even the closer main sequence
stars (within ten parsecs or so), can be resolved and imaged, although
with reduced image fidelity. In these cases, large scale temperature
structures are not expected on the radio photospheres, and the primary
physical parameters of diameter and brightness will be well
constrained.

Using the Hipparcos catalog measurements of stellar spectral type,
luminosity class, and distance for over 100,000 stars, we estimate
that there are at least 10,000 stars that will be resolved by the
ngVLA. While the vast majority of these (95\%) are giants or
supergiants, there are still over 500 main sequence stars that can be
resolved, with $\sim 50$ to 150 of each spectral type (besides O
stars).  Note that these are lower limits, since radio photospheres
can be larger than optical, and the Hipparcos catalog might not be
complete. An initial investigation of the Gaia catalog suggests these
numbers may be pessimistic by a factor of a few.

\vskip 0.1in

\acknowledgements The National Radio Astronomy Observatory is a facility of the National Science Foundation operated under cooperative agreement by Associated Universities, Inc.

\end{document}